\documentclass[11pt,superscriptaddress,aps,prd,
showkeys,notitlepage]{revtex4}

\usepackage{graphicx}
\usepackage{amsmath,amssymb,amsfonts,amstext,latexsym,stmaryrd,amscd,txfonts,pxfonts}
\usepackage{float}

\begin{document}

\title{Energy-momentum and angular-momentum of a gyratonic pp-waves spacetime}

\author{F. L. Carneiro}
\email{fernandolessa45@gmail.com}
\affiliation{Instituto de F\'isica, Universidade de Bras\'ilia, 70910-900, Bras\'ilia, DF, Brazil}

\author{S. C. Ulhoa}
\email{sc.ulhoa@gmail.com}
\affiliation{Instituto de F\'isica, Universidade de Bras\'ilia, 70910-900, Bras\'ilia, DF, Brazil}
\affiliation{International Center of Physics, Instituto de F\'{\i}sica, Universidade de
Bras\'{\i}lia, 70.910-900, Brasilia, DF, Brazil}

\author{J. F. da Rocha-Neto}
\email{rocha@fis.unb.br}
\affiliation{Instituto de F\'isica, Universidade de Bras\'ilia, 70910-900, Bras\'ilia, DF, Brazil}

\begin{abstract}
Gyratonic plane fronted gravitational waves are exact solutions of Einstein's field equations, which correspond to gravitational waves that carry momentum and angular-momentum. Using the definitions of the Hamiltonian formulation of the Teleparallel Equivalent of General Relativity, we explicitly
 evaluate the general expressions of the energy-momentum and angular-momentum of these space-times. In order to better understand the additional properties of these gravitational waves, we consider the motion of particles in this space-time and obtain an interesting relation between the angular-momentum of the particles and that of the gravitational waves.
\end{abstract}
\keywords{Teleparallel gravity; Gyratonic gravitational waves.}

\maketitle

\date{\today}
\section{Introduction}
\par \qquad
The study of plane fronted gravitational waves with parallel rays (pp-waves) are a standard topic in the inspection of exact solutions of Einstein's field equations. Although the existence of these waves are questioned, they represent a valid class of solutions of the gravitational field equations that do not violate any physical principle and are geodesically complete \cite{witten1962gravitation}. The pp-waves space-time were investigated in the 1950's and 1960's specially by Peres, Pirani and Bondi \cite{peres1959some,bondi1959gravitational}, and most recently in the Refs. \cite{zhang2018memory,andrzejewski2018memory,fuster2018berwald,andrzejewski2018niederer,formiga2018energy,zhang2017soft,zhang2018velocity,zhang2018ion}.

The gravitational waves considered in this paper represent the exterior field of spinning particles (gyratons) moving with speed of light. They were quoted by Peres  \cite{peres1959some} and studied by Misner \cite{misner1957classical}. Most recently, the gyratonic waves were rediscovered by Frolov and collaborators \cite{frolov2005gravitational,frolov2005gravitational2}, and studied in great detail in the Refs. \cite{podolsky2014gyratonic,maluf2018kinetic}, where in the latter a direct interaction between the energy of the gravitational field and the kinetic energy of a particle, which is hit by the gyratonic wave, was obtained. Despite the long history of pp-waves research, there are still properties not yet appreciated.

A recent conjecture about the local exchange of energy between particles and pp-waves \cite{maluf2018variations} provides an interesting opportunity to better understand these waves in a general way (gyratonic) by explicitly calculating its energy and angular-momentum. In order tho achieve this, we use the well established expressions that arise in the Hamiltonian formulation of the Teleparallel Equivalent of General Relativity (TEGR). 

Our aim in this paper is to generalize the non-gyratonic pp-waves expressions of the energy-momentum of the Ref. \cite{maluf2008energy} and the angular-momentum of the Ref. \cite{da2014angular}. In order to better understand the gyratonic effect of the gravitational field, we consider two solutions of the Einstein's equations: one axially symmetric similar to the Aichelburg-Sexl monopole solution \cite{aichelburg1971gravitational}, and another which is the monopole solution with the dipole correction.

This article is organized as follows. In section $2$ we briefly present the structure of the TEGR and its equivalence with general relativity, and then we present the definitions of energy-momentum and angular-momentum that arise from the Hamiltonian formulation of the TEGR. In section $3$ the gyratonic metric and the Einstein's vacuum field equations for the gyratonic waves are presented, and in addition a simple relationship between the two functions describing the gyratonic pp-waves is presented. In section $4$, a set of tetrads adapted to a spatially static observer and associated with the gyratonic metric is constructed. Moreover, the energy density of the gyratonic space-time is explicitly evaluated for an asymptotically flat solution. In section $5$, using the tetrads obtained in the previous section, the angular-momentum of the gyratonic field is calculated and compared to that of the non-gyratonic pp-waves. Also in the section $5$, the gyratonic angular-momentum for an axially symmetric solution is calculated. Finally, in section $6$, we present our conclusions and consider the effects of the results on the angular-momentum of a test particle, we also present an interesting relationship between the asymptotic behavior of the angular-momentum of particles and the angular-momentum of the wave. 

We use the following notation: space-time indices are denoted by Greek letters $\mu$, $\nu$, ... and $SO$(3,1) indices are indicated by Latin letters $a$, $b$, ..., which run from 0 to 3. Time and space indices are indicated as $\mu = 0,i$ and $a = (0), (i)$ . The tetrad field is indicated by $e^{a}\,_{\mu}$ and the flat Minkowski space-time metric tensor $\eta_{ab}=diag(-1,1,1,1)$ raises and lowers the Lorentz indices, while the metric tensor $g_{\mu\nu}$ raises and lowers the space-time indices. We use the geometrized units system, i.e., $G=c=1$.

\section{The Teleparallel Equivalent of General Relativity}
\par \qquad
The TEGR is an alternative description, dynamically equivalent to general relativity, constructed in terms of the tetrad field $e^{a}\,_{\mu}$.
The tetrads are reference frames adapted to preferred observers in space-time. The components $e_{(0)}\,^{\mu}$ are always tangent to the observer world line. In this case, the $e_{(0)}\,^{\mu}$  component is identified with the four-velocity of the observer $U^{\mu}$ in their own rest frame. To perform a measurement without the interference of the frame motion, the spatial velocity $U^{i}$ must be zero, i.e., the observer moves along his own world line only. A set of tetrads adapted to a spatially static observer must satisfy the condition
\begin{equation}\label{eq1}
U^{i}=e_{(0)}\,^{i}=0\,.
\end{equation}
The TEGR is constructed in terms of a quadratic combination in the torsion tensor which is related 
to the antisymmetric part of the Cartan connection 
\begin{equation}\label{eq2}
\Gamma^{\mu}\,_{\lambda\nu}=e^{a\mu}\,\partial_{\lambda}e_{a\nu}\,.
\end{equation}
The above connection is not symmetric in the permutation of the lower indices. The Cartan connection is curvature free, but has a non null torsion tensor
\begin{equation}\label{eq3}
T_{a\mu\nu}=\partial_{\mu}e_{a\nu}-\partial_{\nu}e_{a\mu}\,.
\end{equation}
With the torsion tensor above it is possible to obtain a curvature scalar $R(e)$ such that
$$eR(e) = -e\left(\frac{1}{4}T^{abc}T_{abc}+\frac{1}{2}T^{abc}T_{bac}-T^{a}T_{a}\right)+2\partial_{\mu}\left(eT^{\mu}\right)\,,$$
and the Lagrangian density $L$ for the gravitational and matter fields may be written as \cite{maluf2013teleparallel}
\begin{equation}\label{eq4}
L=-ke\Sigma^{abc}T_{abc} - L_{M}\,,
\end{equation}
where
\begin{equation}\label{eq5}
\Sigma^{abc}\equiv\frac{1}{4}\left(T^{abc}+T^{bac}-T^{cab}\right)+\frac{1}{2}\left(\eta^{ac}T^{b}-\eta^{ab}T^{c}\right)\,,
\end{equation}
with $T^{a}\equiv T^{b}\,_{b}\,^{a}$, $e\equiv det(e^{a}\,_{\mu})$, $k = 1/16\pi$ and $L_{m}$ is the Lagrangian density for the matter fields. The field equations are obtained varying the above Lagrangian density with respect to the tetrad field $e_{a\mu}$, thus they read \cite{maluf2013teleparallel}
\begin{equation}\label{eq6}
e_{a\lambda}e_{b\lambda}\partial_{\nu}\left(e\Sigma^{b\lambda\nu}\right)-e\left(\Sigma^{b\nu}\,_{a}T_{b\nu\mu}-\frac{1}{4}e_{a\mu}T_{bcd}\Sigma^{bcd}\right)=\frac{1}{4k}eT_{a\mu}\,,
\end{equation}
where $eT_{a\mu}\equiv\frac{\delta L_{M}}{\delta e^{a\mu}}$ is the projected energt-momentum tensor due the matter fields,.
 
Although the field equations (\ref{eq6}) are dynamically equivalent to Einstein's field equations \cite{maluf2013teleparallel}, their symmetries are not. The absence of the divergence term on the right-hand side of equation (\ref{eq4}) makes $L$ invariant only under global $SO(3,1)$ transformations.
 In order to obtain the Hamiltonian density of the TEGR for the gravitational field ($L_{M}$ = 0) , we rewrite the Lagrangian density $L$ in the phase space as $L=\Pi^{ai}\dot{e}_{ai}-H$, where $\Pi^{ai}=-4k\Sigma^{a0i}$ are the momenta canonically conjugated to $e_{ai}$ and the dot represents the derivative with respect to the time $t$. The Hamiltonian density may then be written as \cite{da2010hamiltonian}
\begin{equation}\label{eq7}
H(e,\Pi)=e_{a0}C^{a}+\lambda_{ab}\Gamma^{ab}\,,
\end{equation}
where $\lambda_{ab}$ and $e_{a0}$ are Lagrange multipliers.

The constraints $C^{a}$ and $\Gamma^{ab}$ in the above Hamiltonian density are first class constraints and are functions of $\Pi^{ai}$ and $e_{ai}$. The constraint $C^{a}$ may be written as
\begin{equation}\label{eq8}
C^{a}=-\partial_{i}\Pi^{ai}+h^{a}=0\,,
\end{equation}
where $h^{a}$ is a very long expression of the field variables (explicitly written in the Ref. \cite{maluf2013teleparallel}). The constraint $\Gamma^{ab}$ is given by
\begin{equation}\label{eq9}
\Gamma^{ab}\equiv 2\Pi^{\left[ab\right]}+4ke\left(\Sigma^{a0b}-\Sigma^{b0a}\right)=2\Pi^{\left[ab\right]}-2k\partial_{i}\left[e\left(e^{ai}e^{b0}-e^{bi}e^{a0}\right)\right]=0\,.
\end{equation}
The constraints above satisfy the algebra of the Poincar\'e group \cite{da2010hamiltonian}.
Both constraints $C^{a}$ and $\Gamma^{ab}$ contain a total divergence and under integration yield expressions for the gravitational energy-momentum and angular-momentum, respectively.

From the integration of the constraint $C^{a}$ in (\ref{eq8}) it is possible to define the energy-momentum four-vector $P^{a}$ as
$$
P^{a}=\int_{V}{h^{a}d^{3}x}=\int_{V}{\partial_{i}\Pi^{ai}d^{3}x}\,.
$$
Since the expression for $h^{a}$ is too long, it is more convenient to work with the right hand side of the above expression. 
The energy-momentum four-vector $P^{a}$ of the gravitational and matter fields, contained in a volume $V$ of space, is then defined as
\begin{equation}\label{eq10}
P^{a}=4k\int_{V}{\partial_{i}\left(e\Sigma^{a0i}\right)d^{3}x}\,.
\end{equation}
The expression (\ref{eq9}) is identically zero, and in analogy to the definitions of the energy-momentum four-vector, the primary constraint $\Gamma^{ab}=0$ under integration give us the angular-momentum of the gravitational field as 
\begin{equation}\label{eq11}
L^{ab}=\int_{V}{M^{ab}d^{3}x}=-2k\int_{V}{\partial_{i}\left[e\left(e^{ai}e^{b0}-e^{bi}e^{a0}\right)\right]d^{3}x}\,,
\end{equation}
where $M^{ab}$ is the gravitational angular-momentum density and $V$ is the three-dimensional volume of the space of interest.

The expressions (\ref{eq10}) and (\ref{eq11}) are both invariant under spatial coordinate transformations and time re-parametrizations, but they are not under local Lorentz transformations. The former make these quantities frame dependent, as happens in classical physics, e.g., a moving observer with velocity $v\neq0$ measures a different energy than a spatially static observer ($v=0$). In the case of a vacuum solution, like the pp-waves, the expressions $(\ref{eq10})$ and $(\ref{eq11})$ represent the energy-momentum four-vector and angular-momentum of the gravitational field, respectively. 

As mentioned previously, the constraints $C^{a}$ and $\Gamma^{ab}$ satisfy the algebra of the Poincar\'e group \cite{da2010hamiltonian}
\begin{eqnarray}
\lbrace C^a , C^b \rbrace &=& 0\,, \nonumber \\
\lbrace C^a , \Gamma^{bc} \rbrace &=& \eta^{ab} C^c-\eta^{ac} C^b 
\,, \nonumber \\
\lbrace \Gamma^{ab}, \Gamma^{cd} \rbrace &=&
\eta^{ac}\Gamma^{bd} +\eta^{bd}\Gamma^{ac} -\eta^{ad}L^{bc}-\eta^{bc}\Gamma^{ad}\,.
\nonumber
\end{eqnarray}
Therefore, the interpretations of $P^{a}$ and $L^{ab}$, which also satisfy the same algebra, are physically consistent.
We shall use these definitions to explicitly evaluate the energy and angular-momentum of the gyratonic space-time, to be presented in the next section.

\section{Gyratonic space-time}
\par \qquad
The gyratonic pp-waves line element is described in terms of generalized Brinkmann coordinates as \cite{podolsky2014gyratonic}
\begin{equation}\label{eq12}
ds^{2}=H(u,\rho,\phi)du^{2}+d\rho^{2}+\rho^{2}d\phi^{2} -2J(u,\rho,\phi)dud\phi + 2dudv\,.
\end{equation}
This wave moves with the speed of light, so the wave front is always at $u=0$ and the surfaces $u=constant$ are flat.
The above line element represents the field generated by spinning particles that move at the speed of light, named ``gyratonic'' by Frolov and Fursaev \cite{frolov2005gravitational}. Outside the source, this metric represents a pure radiation field that propagates along the null direction $v$. In the asymptotic limit, one may identify
\begin{equation}\label{eq13}
u\equiv\frac{z-t}{\sqrt{2}}
\end{equation}
and
\begin{equation}\label{eq14}
v\equiv\frac{z+t}{\sqrt{2}}\,,
\end{equation}
where the $z$ axis represents the propagation axis of the wave.

The gyratonic metric is specified by two, in principle, independent functions $H(u,\rho,\phi)$ and $J(u,\rho,\phi)$. These functions must be periodic in the angular coordinate $\phi$, and if they are independent of $\phi$, the space-time is axially symmetric around the propagation axis. If the functions $H$ and $J$ are written as 
\begin{equation}\label{eq15}
H=\omega^{2} (u)\rho^{2}+2\omega (u)\chi (u,\phi)+H_{0}(u,\rho,\phi)
\end{equation}
and
\begin{equation}\label{eq16}
J=\omega (u)\rho^{2}+\chi (u,\phi)\,,
\end{equation}
respectively, the Einstein's vacuum field equations are reduced to \cite{podolsky2014gyratonic}
\begin{equation}\label{eq17}
\nabla^{2}H_{0}\equiv \partial_{\rho}\partial_{\rho}H_{0}+\frac{1}{\rho}\partial_{\rho}H_{0}+\frac{1}{\rho^{2}}\partial_{\phi}\partial_{\phi}H_{0}=\frac{2}{\rho^{2}}\left(\partial_{u}\partial_{\phi}\chi-\omega\partial_{\phi}\partial_{\phi}\chi\right)\,.
\end{equation}
As mentioned in the Ref. \cite{podolsky2014gyratonic}, there exists a gauge freedom $f(u)$ in the choice of the angular coordinate, resulting in the gauge $\tilde{\omega}(u)=\omega(u)+\partial_{u}f(u)$. The function $\omega(u)$ in the equation (\ref{eq15}) may be set to zero by an appropriate gauge transformation. With this simplification, the gyratonic function $J$ becomes
\begin{equation}\label{eq18}
J=J(u,\phi)=\chi(u,\phi)
\end{equation}
and the equation (\ref{eq17}) is simplified to
\begin{equation}\label{eq19}
\nabla^{2}H_{0}=\frac{2}{\rho^{2}}\left(\partial_{u}\partial_{\phi}J\right)\,.
\end{equation}

Applying separation of variables in the functions $J$ and $H$ as
$$
J(u,\phi)=j_{1}(u)j_{2}(\phi)
$$
and
$$
H(u,\rho,\phi)=h_{1}(u)h_{2}(\rho,\phi)\,,
$$
the equation (\ref{eq19}) becomes
\begin{equation}\label{eq20}
\frac{\rho^{2}\nabla^{2}h_{2}(\rho,\phi)}{2\partial_{\phi}j_{2}(\phi)}=\frac{\partial_{u}j_{1}(u)}{h_{1}(u)}=constant\equiv\vartheta\,.
\end{equation}
Thus, we have two different equations, namely
\begin{equation}\label{eq21}
\rho^{2}\partial_{\rho}\partial_{\rho}h_{2}+\rho\partial_{\rho}h_{2}+\partial_{\phi}\partial_{\phi}h_{2}=2\vartheta\partial_{\phi}j_{2}(\phi)
\end{equation}
and
\begin{equation}\label{eq22}
\partial_{u}j_{1}(u)=\vartheta h_{1}(u)\,.
\end{equation}
A particular solution for $h_{2p}$, satisfying the equation (\ref{eq21}), with the homogeneous equation $\nabla^{2}h_{2h}=0$, is given by
$$
h_{2p}(\rho,\phi)=\alpha\ln{\rho}+2\vartheta f(\phi)\,,
$$
where $\partial_{\phi}f(\phi)=j_{2}(\phi)$ and $\alpha=constant$.
It should be noted that equation (\ref{eq22}) allows an explicit relation between $H$ and $J$ on the variable $u$ . Also from the expression (\ref{eq18}), if $\chi=\chi(u)$, then $H$ must satisfy $\nabla^{2}H=0$. In addition to an arbitrary dependence in $u$, the metric functions can be explicitly determined by the Einstein's equations. This fact is typical of waves solutions, where the geometry of the wave pulse may be chosen.

In the next two sections, we evaluate the quantities (\ref{eq10}) and (\ref{eq11}) for the gyratonic space-time.
In this point, it is important to emphasize that in what follows we will consider solutions of Einstein's equations only in the vacuum 
regions outside the gyratonic matter source, i.e., in regions where $\rho > \rho_{0}$ and the energy-momentum tensor vanishes ($T_{\mu\nu} = 0$). In fact, for a more realistic analysis of these solutions it would be necessary to know the solutions inside the gyratonic matter source, i.e., in regions where $\rho < \rho_{0}$ and the energy-momentum tensor is such that $T_{\mu\nu} \neq 0$. In addition, as mentioned in the Ref.\cite{PRD75}, for physically realistic solutions it is expected that the gyratonic matter source has finite radius $\rho = \rho_{0} \neq 0$.
\section{The energy-momentum of gyratonic PP-Waves}
\par \qquad
For the evaluation of the energy-momentum of a gravitational field in the TEGR, we need of a set of tetrads $e_{a\mu}$ associated with the space-time and adapted to a spatially static observer. The energy expression (\ref{eq10}) comes from a secondary constraint of the Hamiltonian formulation, so the metric must be written in the $(t,\rho,\phi,z)$ coordinates using the expressions (\ref{eq13}) and (\ref{eq14}). In these coordinates the metric (\ref{eq12}) reads
\begin{equation}\label{eq23}
ds^{2}=\left(\frac{H}{2}-1\right)dt^{2}+d\rho^{2}+\rho^{2}d\phi^{2}+\sqrt{2}Jdtd\phi-\sqrt{2}Jdzd\phi+\left(1+\frac{H}{2}\right)dz^{2}-Hdtdz\,.
\end{equation}

The frame is determined by fixing six conditions in the tetrad field. The fact that
\begin{equation}\label{eq24}
U^{\mu}=e_{(0)}\,^{\mu}=\left(\frac{1}{A},0,0,0\right)
\end{equation}
 ensures that the observers adapted to this set of tetrads follows a time like world line and in view of equation (\ref{eq1}) the observers do not have any spatial translation movement.
The four-velocity $U^{\mu}$ is a time like vector, i.e., $U^{2} = U^{\mu}U_{\mu} = g_{00}/A^{2} = -1$ and this result is independent of values of $H$, however to evaluate the physical quantities we emphasize that the tetrad field is valid in space time regions where $2 - H > 0$. Anyway, it should be expected that in regions far way from the source these waves have small amplitudes.

The others conditions fix the spatial orientation of the frame, i.e., $e_{(1)}\,^{\mu}$, $e_{(2)}\,^{\mu}$ and $e_{(3)}\,^{\mu}$ are asymptotically unit four-vectors along the directions of $x$, $y$ and $z$, respectively.
A suitable set of tetrads adapted to a spatially static observer and associated with the line element (\ref{eq23}) is given by
\begin{equation}\label{eq25}
e_{a\mu}=\left(
\begin{array}{cccc}
 -A& 0 & \frac{J}{\sqrt{2}A} & -B \\
 0 & \cos (\phi ) & -\rho  \sin (\phi ) & 0 \\
 0 & \sin (\phi ) & \rho  \cos (\phi ) & 0 \\
 0 & 0 & -\frac{J}{\sqrt{2}A} & C \\
\end{array}
\right)\,,
\end{equation}
were $A=\sqrt{1-H/2}$, $B=\frac{H}{2\sqrt{1-H/2}}$, $C=\frac{1}{\sqrt{1-H/2}}$ and $e=det(e^{a}\,_{\mu})=\rho$ is the determinant of the tetrad field.
The set of inverse tetrads may be obtained by the relation $e^{a\mu}=\eta^{ab}g^{\mu\nu}e_{b\nu}$ and it reads
\begin{equation}\label{eq26}
e^{a\mu}=\left(
\begin{array}{cccc}
 -1/A& 0 &0 & 0 \\
 -\frac{J\sin{\phi}}{\sqrt{2}\rho} & \cos (\phi ) & -\frac{\sin (\phi )}{\rho} & -\frac{J\sin{\phi}}{\sqrt{2}\rho} \\
 \frac{J\cos{\phi}}{\sqrt{2}\rho} & \sin (\phi ) & \frac{\cos (\phi )}{\rho} & \frac{J\cos{\phi}}{\sqrt{2}\rho} \\
- \frac{H}{\sqrt{2}A} & 0 & 0 & A \\
\end{array}
\right)\,.
\end{equation}
It is possible to see that by putting $H=0=J$, one obtains $e_{a}\,^{\mu}=\delta_{a}^{\mu}$ and the torsion $T_{a\mu\nu}$ vanishes, which excludes the necessity to regularize the field.

With the set of tetrads (\ref{eq25}), the energy-momentum of the gravitational field can be obtained. In order to achieve such aim, first we calculate the non-vanishing components of $T^{abc}=e^{b\mu}e^{c\nu}T^{a}\,_{\mu\nu}$, reading
\begin{equation}\nonumber
\begin{split}
&T^{(0)(0)(1)}=-T^{(3)(1)(3)}=-\frac{1}{4\rho A^{2}}\left(2\sqrt{2}\sin{\phi}\partial_{t}J-\sin{\phi}\partial_{\phi}H+\rho\cos{\phi}\partial_{\rho}H\right)\,,\\
&T^{(0)(0)(2)}=-T^{(3)(2)(3)}=-\frac{1}{4\rho A^{2}}\left(-2\sqrt{2}\cos{\phi}\partial_{t}J+\cos{\phi}\partial_{\phi}H+\rho\sin{\phi}\partial_{\rho}H\right)\,,\\
&T^{(0)(0)(3)}=T^{(3)(0)(3)}=-\frac{1}{4A^{3}}\partial_{t}H\,,\\
&T^{(0)(1)(3)}=\frac{1}{2\rho A^{2}}\left(\sqrt{2}\sin{\phi}\partial_{t}J-\sin{\phi}\partial_{\phi}H+\rho\cos{\phi}\partial_{\rho}H\right)\,,\\
&T^{(0)(2)(3)}=\frac{1}{2\rho A^{2}}\left(-\sqrt{2}\cos{\phi}\partial_{t}J+\cos{\phi}\partial_{\phi}H+\rho\sin{\phi}\partial_{\rho}H\right)\,,\\
&T^{(3)(0)(1)}=-\frac{1}{\sqrt{2}\rho A^{2}}\sin{\phi}\partial_{t}J\,,\\
&T^{(3)(0)(2)}=\frac{1}{\sqrt{2}\rho A^{2}}\cos{\phi}\partial_{t}J\,,
\end{split}
\end{equation}
where we made use of $(\partial_{t}+\partial_{z})(H,J)=0$. From the above expressions and the definition in (\ref{eq5}), the non-null components $\Sigma^{a0\nu}$ are
\begin{equation}\label{eq27}
\begin{split}
&\Sigma^{(0)01}=\Sigma^{(3)01}=-\frac{1}{4\sqrt{2}}\frac{\partial_{\rho}H}{\sqrt{2-H}}\\
&\Sigma^{(0)02}=\Sigma^{(3)02}=-\frac{1}{4\sqrt{2}\rho^{2}}\frac{\partial_{\phi}H}{\sqrt{2-H}}+\frac{\partial_{t}J}{2\rho^{2}\sqrt{2-H}}\\
&\Sigma^{(1)01}=\frac{1}{4}\frac{\partial_{t}H}{2-H}\cos{\phi}\\
&\Sigma^{(1)02}=-\frac{1}{4\rho}\frac{\partial_{t}H}{2-H}\sin{\phi}\\
&\Sigma^{(1)03}=-\frac{1}{4\rho}\frac{\partial_{\phi}H\sin{\phi}-\rho\partial_{\rho}H\cos{\phi}}{2-H}\\
&\Sigma^{(2)01}=\frac{1}{4}\frac{\partial_{t}H}{2-H}\sin{\phi}\\
&\Sigma^{(2)02}=\frac{1}{4\rho}\frac{\partial_{t}H}{2-H}\cos{\phi}\\
&\Sigma^{(2)03}=\frac{1}{4\rho}\frac{\partial_{\phi}H\cos{\phi}+\rho\partial_{\rho}H\sin{\phi}}{2-H}\,.
\end{split}
\end{equation}
Finally, from equation (\ref{eq10}) and the expressions in (\ref{eq27}) we have
\begin{equation}
P^{(0)}=P^{(3)}=-\frac{k}{8}\int_{V}{d^{3}x\left[\partial_{\rho}\left(e\Sigma^{(0)01}\right)+[\partial_{\phi}\left(e\Sigma^{(0)02}\right)\right]}\,.
\end{equation}
Using the equation (\ref{eq19})  and noticing that $\partial_{t}J=-\frac{1}{\sqrt{2}}\partial_{u}J$, the above expression reduces to
\begin{equation}\label{eq29}
P^{(0)}=P^{(3)}
=-k\int_{V}{\left(\frac{\rho^{2}(\partial_{\rho}H)^2 + (\partial_{\phi}H)^2 + 2\partial_{u}J\partial_{\phi}H}{8\rho A^{3}}+2\frac{\partial_{u}\partial_{\phi}J}{\rho A}\right)d^{3}x}\,.
\end{equation}
The remaining components of the energy-momentum four-vector vanish, i.e.,
\begin{equation}\label{eq30}
P^{(1)}=0=P^{(2)}\,.
\end{equation}
The square of the energy-momentum four-vector is null, as it happens for the non gyratonic pp-wave, i.e., $P^{a}P_{a}=0$ \cite{maluf2008energy}. This result is consistent with the fact that these fields describe massless particles.

In the expression (\ref{eq29}), the non gyratonic pp-waves space-time may be obtained only by taking $J=0$. This can be seen by comparing the results obtained here with those obtained in the Ref. \cite{maluf2008energy} by identifying $x=\rho\cos{\phi}$ and $y=\rho\sin{\phi}$. This shows that the gyratonic pp-waves are more general than the non gyratonic pp-waves.

The result in equation (\ref{eq29}) has two interesting features. First, if the gravitational field is axially symmetric, i.e., $H=H(u,\rho)$ and $J=J(u)$, the energy of the wave will be the same of the non gyratonic case. The effect of the gyratonic term in the gravitational energy may only be detected in waves that are not axially symmetric so, for these solutions, the gyratonic wave cannot be distinguished from a non-gyratonic wave by its gravitational energy. Second, the non gyratonic pp-waves have only negative energy \cite{maluf2008energy}, but the gyratonic waves may have positive energy.

\subsection{Multipole solution}
\par \qquad

\begin{figure}[htbp]
	\centering
		\includegraphics[width=0.60\textwidth]{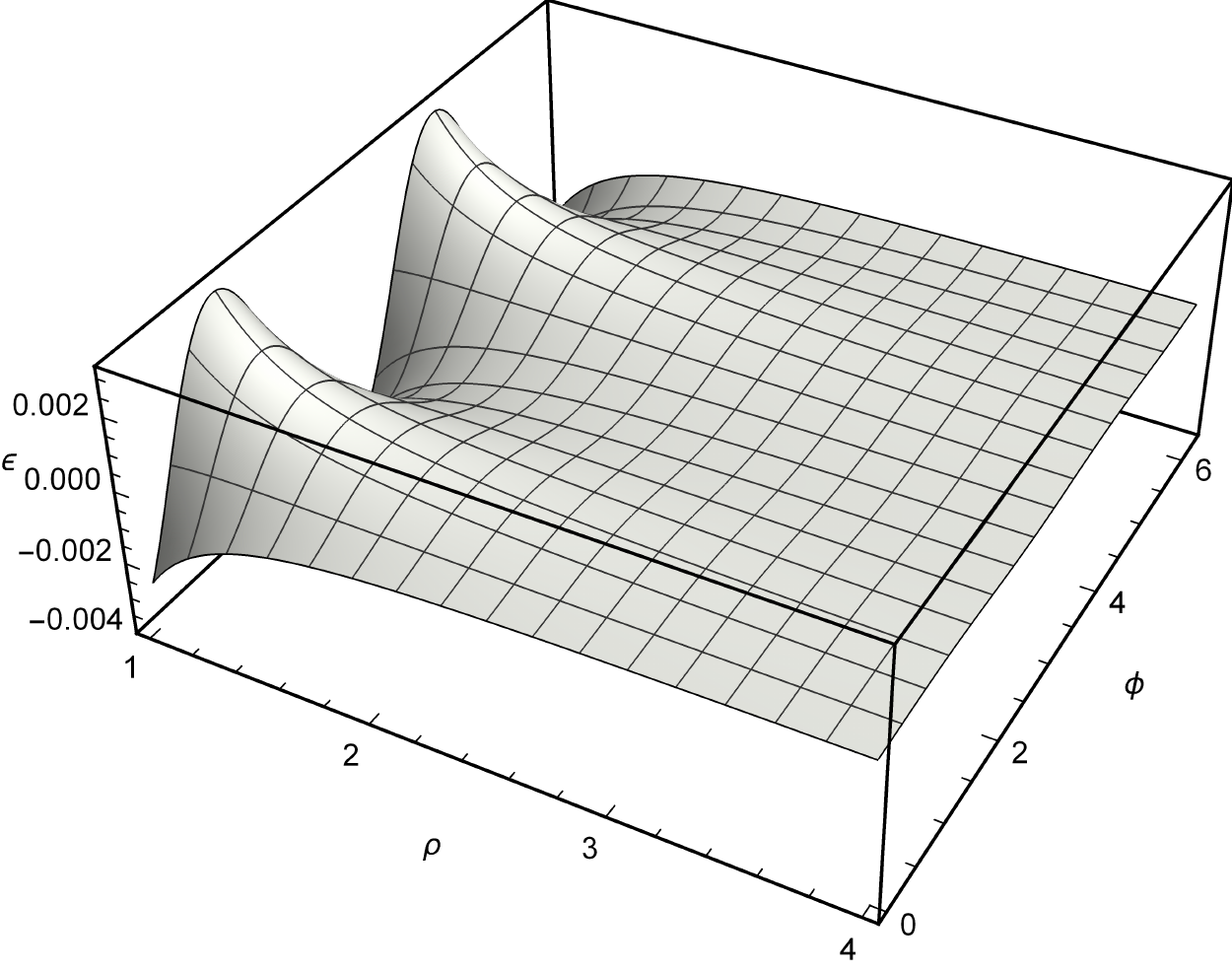}
		\caption{Energy density (\ref{eq33}) on the wave front $u=0$.}
	\label{fig1}
\end{figure}

\begin{figure}[htbp]
\centering
\begin{minipage}[t]{.49\textwidth}
\centering
	\includegraphics[width=1\textwidth]{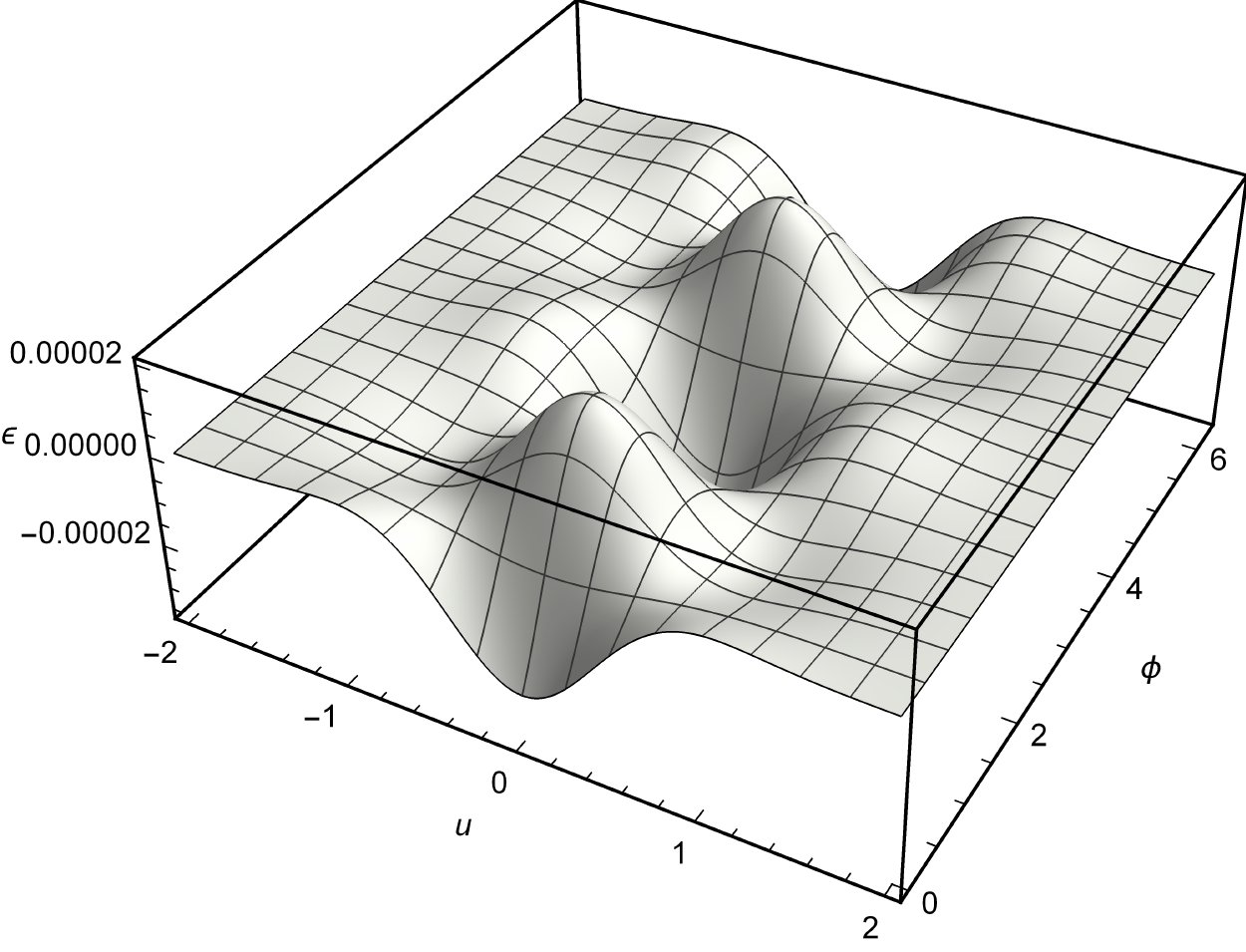}
	\caption{Energy density (\ref{eq33}) for $\rho=5$.}
	\label{fig2}
\end{minipage}
\begin{minipage}[t]{.49\textwidth}
\centering
	\includegraphics[width=1\textwidth]{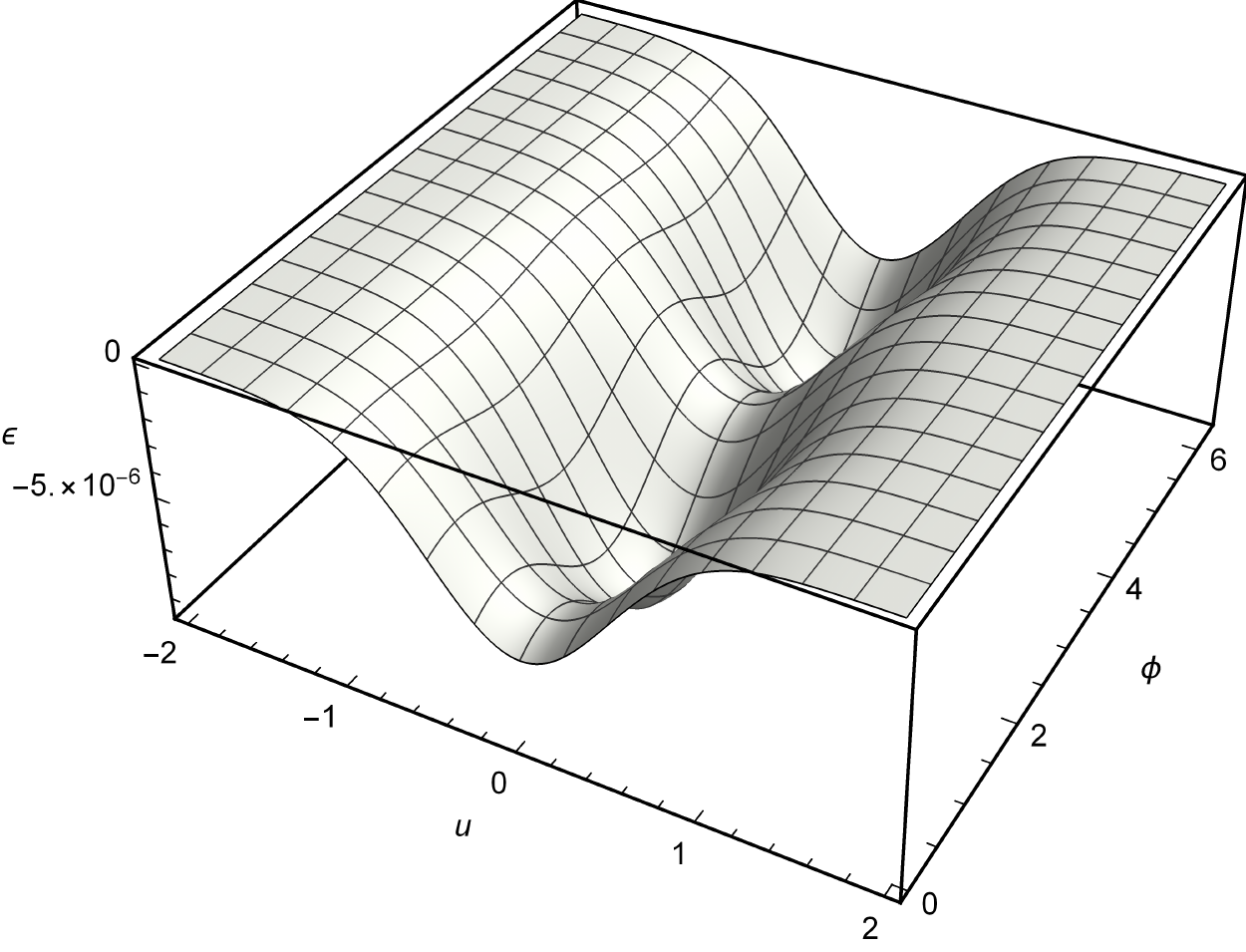}
	\caption{Energy density (\ref{eq33}) for $\rho=5$ and $J=0$.}
	\label{fig3}
\end{minipage}
\end{figure}

In order to better understand the effects of the gyratonic term on the pp-waves, in this subsection the expression (\ref{eq29}) is evaluated for a multipole solution for $H$ with the function $J$ depending only on $u$, i.e., $J=J(u)$ 
and in this case $j_{1}(u)$ and $h_{1}(u)$ are arbitrary functions.
For $H$ we choose the monopole solution with the dipole correction, namely,
\begin{equation}\label{eq31}
\begin{split}
H
&=-\left\{\frac{1}{8}\ln{\sqrt{x^{2}+y^{2}}}-\frac{1}{4}\left[\frac{xy}{\left(x^{2}+y^{2}\right)^{2}}+\frac{x^2-y^2}{\left(x^{2}+y^{2}\right)^{2}}\right]\right\}e^{-u^{2}}\\
&=-\frac{1}{8}\left[\ln\rho+\frac{1}{\rho^{2}}\left(\sin{2\phi}+\cos{2\phi}\right)\right]e^{-u^{2}}\,.
\end{split}
\end{equation}
and we choose
\begin{equation}\label{eq32}
J=\frac{d}{du}e^{-u^2}\,.
\end{equation}
Replacing $H$ and $J$ in the argument of the integral (\ref{eq29}) with the solution (\ref{eq31},\ref{eq32}), we obtain the energy density $\epsilon$ given by
\begin{equation}\label{eq33}
\epsilon\equiv 4k\partial_{i}\left(e\Sigma^{(0)0i}\right)=-\frac{e^{-2 u^2} \left[\rho ^4+4 \rho ^2 \left(32 u^2-17\right) \sin (2 \phi )-4 \rho{^2} \left(32 u^2-15\right) \cos (2 \phi )+8\right]}{8192 \pi  \rho ^5}\,.
\end{equation}
The above energy density depends on three variables, then it is not possible to plot a four-dimensional figure, however, we can plot a contour surface for $u = 0$. The flatness in the asymptotic limit can be seen in Fig. \ref{fig1}
for the wave front at $u=0$.
For a fixed radial position $\rho =5$, the plot of $\epsilon$ is displayed in Fig. \ref{fig2} for the gyratonic wave. It can be compared with the non gyratonic case in Fig. \ref{fig3}. The gravitational energy density is not dependent on the angular variable in the non gyratonic case. For the gyratonic wave, there are regions of positive and negative energy density, depending on the azimuthal coordinate, while in the non gyratonic wave the energy density is always negative.

\begin{figure}[htbp]
\centering
		\includegraphics[width=0.6\textwidth]{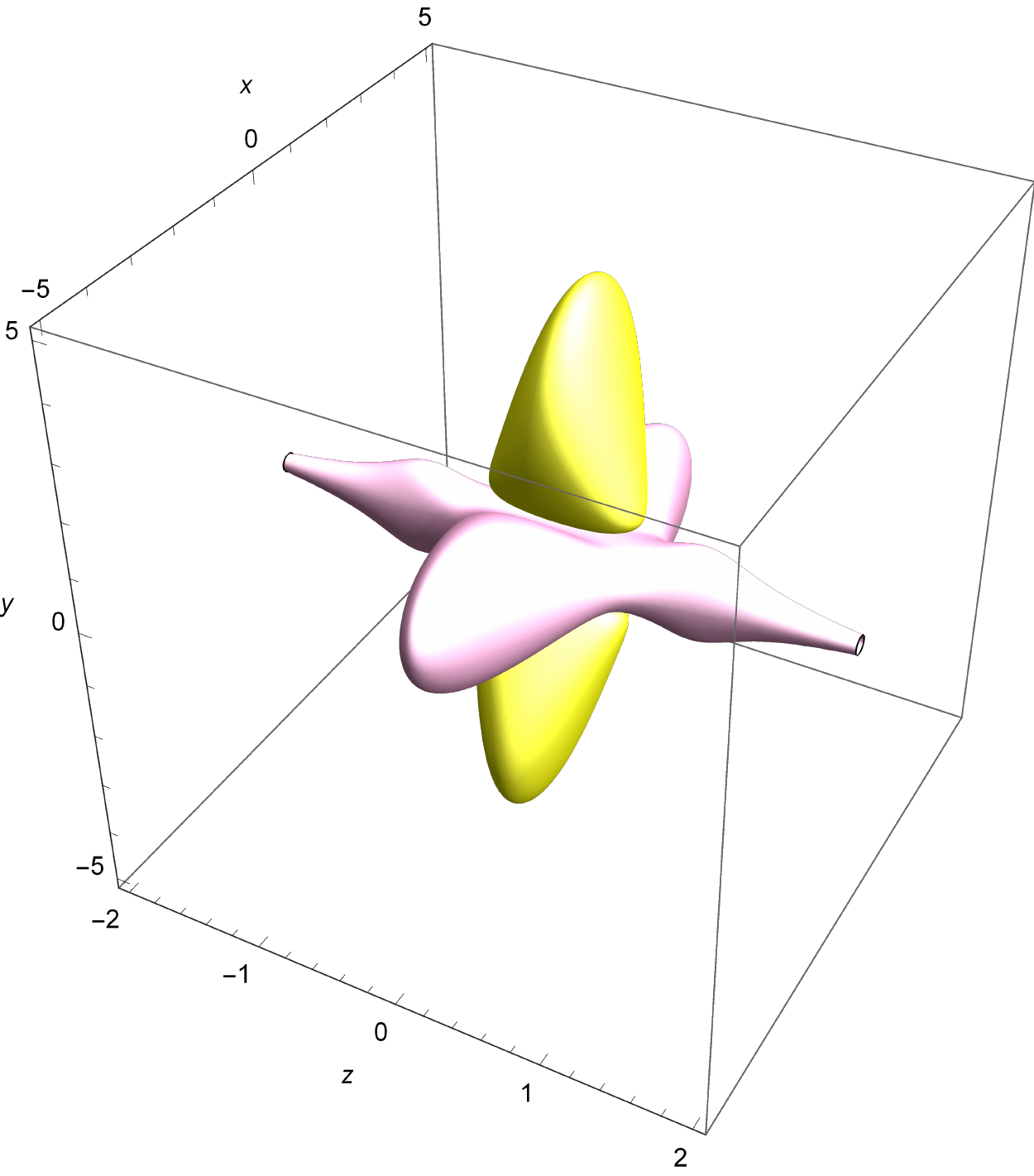}
		\caption{Contour plot for the energy density (\ref{eq33}) at $t=0$.}
	\label{fig4}
\end{figure}

Although the energy density $\epsilon$ depends on three variables, for fixed values of $t$ it is possible to plot levels surfaces for fixed values of $\epsilon$. 
For instance, with the aid of the relation (\ref{eq13}) for $t = 0 \Rightarrow u = z/\sqrt{2}$, the level surface 
can be seen in Fig. \ref{fig4}, where in the pink region the energy density is negative ($\epsilon =-0.0012$) and in the yellow region it is positive ($\epsilon =+0.0012$).

\begin{figure}[htbp]
	\centering
		\includegraphics[width=0.7\textwidth]{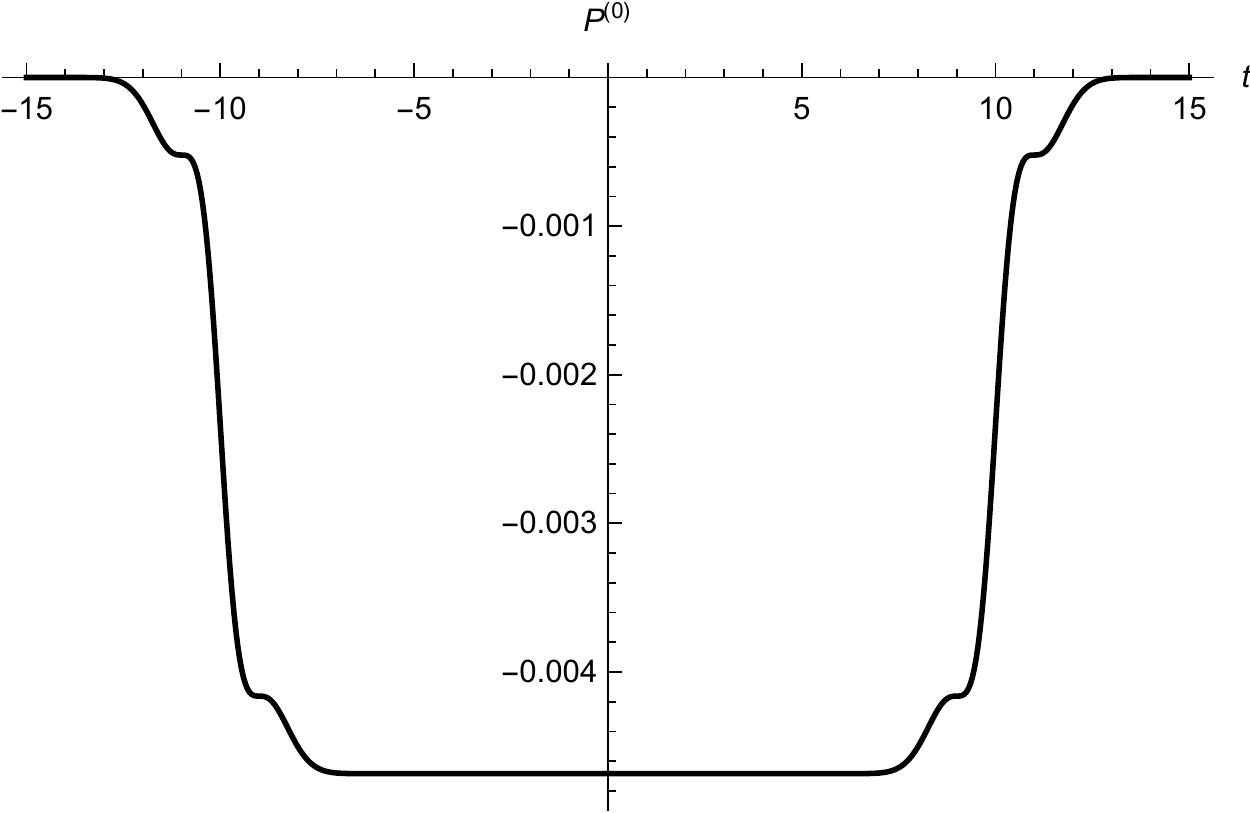}
		\caption{Gravitational energy in the region $20>\rho >1$ as a function of the time $t$.}
	\label{fig5}
\end{figure}

Also with the aid of the relation (\ref{eq13}), the gravitational energy density can be numerically integrated to obtain $P^{0}(t)$ in a sufficiently large $z$ interval, excluding the region around the $z$ axis ($\rho = 0$), this is because this region contains a singularity in the axis $\rho =0$.  The result is presented in Fig. \ref{fig5}. The study of singularities in pp-waves space-times is a recent topic of research  \cite{wang2018singularities}.
The peak in Fig. \ref{fig5} represents the total energy of the gyratonic pp-wave that a spatially static observer will measures. For a particle that interacts with the wave, the difference in the energy of the particle before and after the wave hits the particle, represents the energy absorbed or emitted by the particle. This difference depends on the direction of the acceleration of the gravitational field, as well on the initial conditions of the particle \cite{2018arXiv180809589M}. The peak value in Fig. 5 represents the maximum energy that the particle  can absorb from the wave.
\section{The  angular-momentum of gyratonic PP-Waves}
\par \qquad
In this section, we calculate the components of the angular-momentum of a gyratonic pp-wave. We will use the set of tetrads (\ref{eq26}) that is adapted to a spatially static observer. With the expression (\ref{eq11}) and after some calculations, it is possible to obtain the non-vanishing components of the gravitational angular-momentum, they are
\begin{align}
&L^{(0)(1)}
=-2k\int_{V}{d^{3}x\left[\frac{-\sin{\phi}\partial_{\phi}H+\rho\cos{\phi}\partial_{\rho}H}{\sqrt{2}\left(2-H\right)^{3/2}}-\sin{\phi}\partial_{z}\left(\frac{J}{\sqrt{2-H}}\right)\right]}\,,\label{eq34}\\
&L^{(0)(2)}
=-2k\int_{V}{d^{3}x\left[\frac{\cos{\phi}\partial_{\phi}H+\rho\sin{\phi}\partial_{\rho}H}{\sqrt{2}\left(2-H\right)^{3/2}}+\cos{\phi}\partial_{z}\left(\frac{J}{\sqrt{2-H}}\right)\right]}\,,\label{eq35}\\
&L^{(1)(3)}
=-2k\int_{V}{d^{3}x\left[\left(4-H\right)\frac{\sin{\phi}\partial_{\phi}H-\rho\cos{\phi}\partial_{\rho}H}{2\sqrt{2}\left(2-H\right)^{3/2}}+\sin{\phi}\partial_{z}\left(\frac{J}{\sqrt{2-H}}\right)\right]}\,,\label{eq36}\\
&L^{(2)(3)}
=-2k\int_{V}{d^{3}x\left[-\left(4-H\right)\frac{\cos{\phi}\partial_{\phi}H+\rho\sin{\phi}\partial_{\rho}H}{2\sqrt{2}\left(2-H\right)^{3/2}}-\cos{\phi}\partial_{z}\left(\frac{J}{\sqrt{2-H}}\right)\right]}\,.\label{eq37}
\end{align}
Taking $J=0$, we obtain the same results presented in Ref. \cite{da2014angular} for the case of non gyratonic wave.

The components $L^{(0)(i)}$ are related to the gravitational center of mass \cite{maluf2016teleparallel} and to boosts in the $(i)$ direction. The components $L^{(2)(3)}$ and  $L^{(3)(1)}$ are related to rotations around the $x$ and $y$  axes, respectively. The local indices are always those of the flat space-time, coinciding with the space-time indices on the flat wave front. Therefore, it is possible to identify $M^{(2)(3)}= M_{x}$ and $M^{(1)(3)}= -M_{y}$, so from equations (\ref{eq36}) and (\ref{eq37}), the angular-momentum vector density is given by
\begin{equation}\label{eq38}
\begin{split}
\vec{M}
=&2k\left(M_{x}\hat{x}+M_{y}\hat{y}\right)\\
=&2k\left\{\left[\frac{4-H}{8A^{3}}\partial_{\phi}H+\frac{1}{2}\partial_{u}\left(\frac{J}{A}\right)\right]\hat{\rho}-\left[\frac{4-H}{8A^{3}}\rho\partial_{\rho}H\right]\hat{\phi}\right\}\,,
\end{split}
\end{equation}
where we have made use of $\hat{x}=\hat{\rho}\cos{\phi}-\hat{\phi}\sin{\phi}$ and $\hat{y}=\hat{\rho}\sin{\phi}+\hat{\phi}\cos{\phi}$ in standard cylindrical coordinates.

It is possible to see that an axially symmetric gyratonic space-time has a distinct angular-momentum, i.e., is not the same of the non gyratonic pp-wave. The gyratonic pp-waves carry the rotational character of the source, so these waves are expected to bring some information about the angular-momentum of its source. We note from equation (\ref{eq38}) that when the pp-wave is axially symmetric, the term in the radial component of the angular-momentum density is due to the gyratonic term only. This fact will be explored in the following subsection. Since
for gyratonic and non gyratonic pp-waves with axial symmetry, the energy is the same 
(see equation (\ref{eq29})) and the effect of the gyratonic term in the pp-waves may be analysed only in terms of the angular-momentum. The gyratonic term does not affect the azimuthal angular-momentum of the wave.

\subsection{Axially symmetric solution}

\par
\begin{figure}[htbp]
\centering
\begin{minipage}[t]{.45\textwidth}
\centering
	\includegraphics[width=1.0\textwidth]{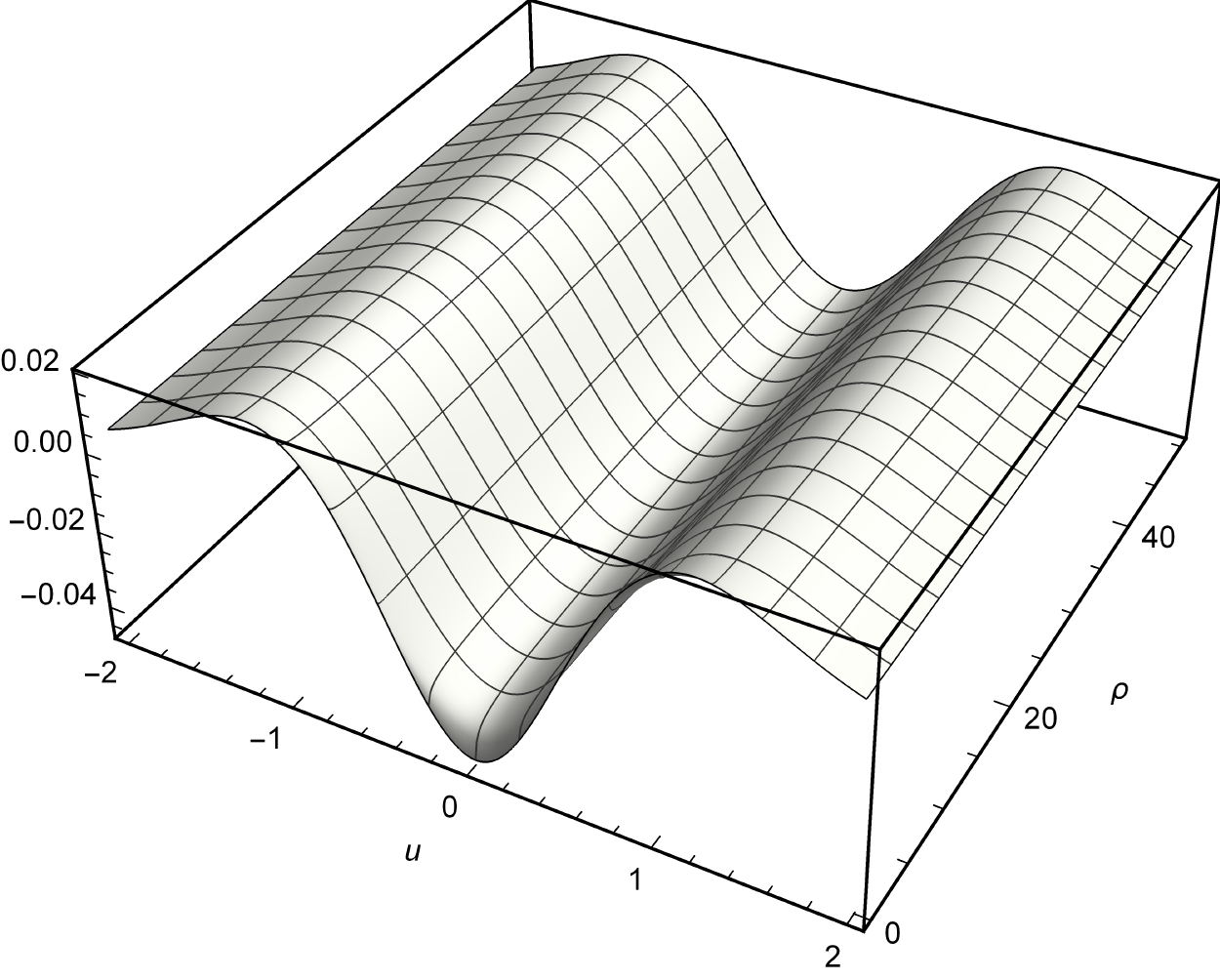}
	\caption{$M_{\rho}$ given by equation (\ref{eq41}).}
	\label{fig6}
\end{minipage}
\begin{minipage}[t]{.45\textwidth}
\centering
	\includegraphics[width=1.0\textwidth]{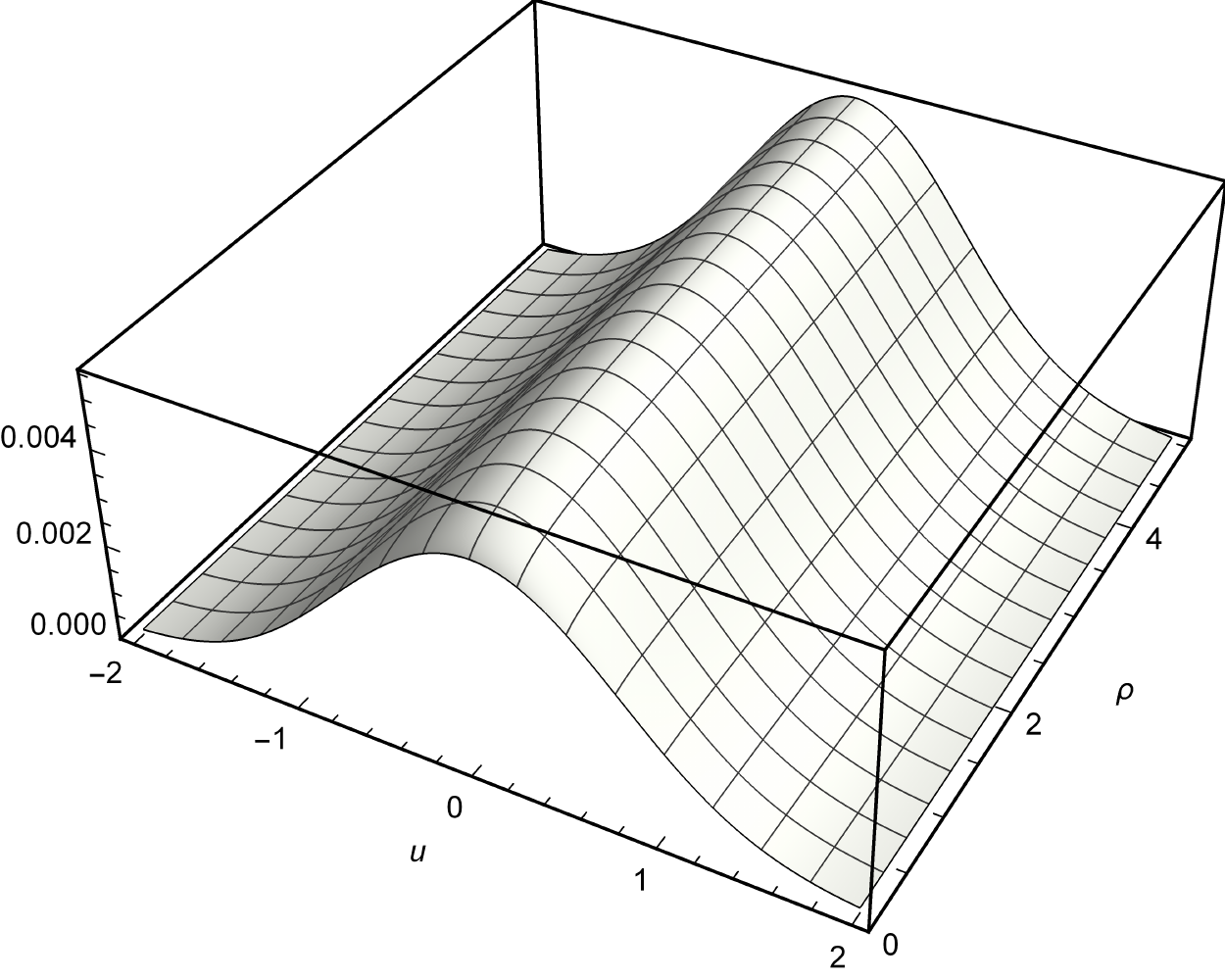}
	\caption{$M_{\phi}$ given by equation (\ref{eq42}).}
	\label{fig7}
\end{minipage}
\end{figure}

 \qquad
In this subsection, the expression (\ref{eq38}) is evaluated for a solution of the equation (\ref{eq19}). The energy of an axially symmetric solution is the same for gyratonic and non gyratonic pp-waves, so the distinction between the two cases is caused solely by the radial angular-momentum of the wave. We choose an Aichelburg-Sexl type solution
\begin{equation}\label{eq39}
H=-\frac{1}{4}\ln{(\rho/\rho_0)}e^{-u^2}\,,
\end{equation}
where in the following we take $\rho_0 =1$, 
and
\begin{equation}\label{eq40}
J=\frac{d}{du}e^{-u^2}\,,
\end{equation}
so the components of the angular-momentum density in (\ref{eq38}) are given by
\begin{equation}\label{eq41}
M_{\rho}=\frac{e^{-2 u^2}}{64 \pi A^{3}} \left[\left(u^2-1\right) \log (\rho )+8 e^{u^2} \left(2 u^2-1\right)\right]
\end{equation}
and
\begin{equation}\label{eq42}
M_{\phi}=\frac{e^{-u^2}}{256 \pi } \left(\frac{1}{4} e^{-u^2} \log (\rho )+4\right)\,.
\end{equation}
Since the above components are independent of the angular variable $\phi$, it is possible to plot them in terms of the variables $\rho$ and $u$. The component $M_{\rho}$ is plotted in Fig. \ref{fig6} and $M_{\phi}$ in Fig. \ref{fig7}. 
From equation (\ref{eq38}) we may notice that in the non gyratonic case, $M_{\rho}$ is null everywhere for $H$ given by (\ref{eq39}) and $M_{\phi}$ has the same expression (\ref{eq42}).

The total gravitational angular-momentum contained in a finite volume $V$, which excludes the axis $\rho=0$, may be obtained performing a numerical integration of the quantities
in equations (\ref{eq36}) and (\ref{eq37}). In terms of the coordinates $x, y$ 
and $z$, this procedure gives
$$\vec{L}=\int_{V}{M_{x}\hat{x}d^{3}x}+\int_{V}{M_{y}\hat{y}d^{3}x}=0\,.$$
The total gravitational angular-momentum vector $\vec{L}$, unlike the gravitational energy $P^{(0)}$, consists of a vector field. Since the space is axially symmetric the positive and negative contributions, in the integration above, cancel each other out. Nevertheless, it is possible to obtain a non-null angular-momentum in a specific region of space, e.g., the integration of the angular-momentum density over the region ($0<\phi <\pi$) yields the exactly opposite of the integration over the region ($\pi<\phi <2\pi$), i.e.,
$$\int_{-L}^{L}dz\int_{\rho_{0}}^{R}d\rho\int_{0}^{\pi}d\phi M_{x,y}=-\int_{-L}^{L}dz\int_{\rho_{0}}^{R}d\rho\int_{\pi}^{2\pi}d\phi M_{x,y}\neq 0 \,.$$
The same happens in the case $J=0$ for an axially symmetric gravitational wave, not being an exclusive gyratonic behavior.

\section{Final considerations}
\par \qquad

In this article we reviewed some important aspects of the TEGR, which is a formulation where the effects of the gravitational field are described in terms of the torsion tensor. In this formalism it is possible to define physical quantities, namely, the energy-momentum four-vector and the angular-momentum of the gravitational field. These physical quantities are invariant under coordinates transformations and time reparametrizations. 
Using these definitions we calculated the energy and the angular-momentum of a gyratonic pp-wave, which was presented in the section $3$ of this work.
The energy of the gyratonic wave yields the expression (\ref{eq29}), which is reduced to the 
case of non gyratonic pp-wave when the function $J=0$. Therefore, the gyratonic pp-wave is a generalization of the non gyratonic pp-wave. The fact that the energy of a gyratonic pp-wave is  distinct of the non gyratonic case and the fact that the definition of energy in equation (\ref{eq10}) is coordinate independent, goes towards the argument presented in Ref. \cite{podolsky2014gyratonic} on the loss of properties of the gravitational field when the term $J$ is not considered in the pp-waves.
The total energy for a gyratonic space-time was integrated using  the expression (\ref{eq33}), and the result was presented in Fig. \ref{fig5}. The way in which the energy of the gravitational wave is altered when it interacts with a particle is yet to be determined.

\begin{figure}[htbp]
\centering
\begin{minipage}[t]{.45\textwidth}
	\includegraphics[width=1\textwidth]{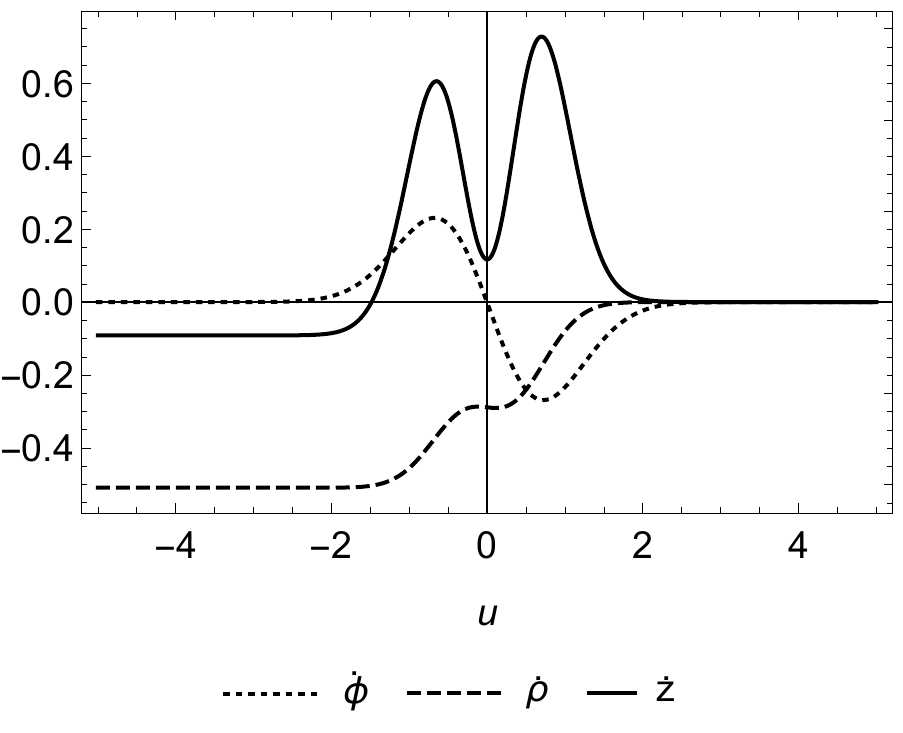}
	\caption{Velocity of a free particle initially at rest in $\rho (u\rightarrow\infty)=5$ with $J$ given by (\ref{eq43}).}
	\label{fig8}
\end{minipage}
\begin{minipage}[t]{.45\textwidth}
	\includegraphics[width=1\textwidth]{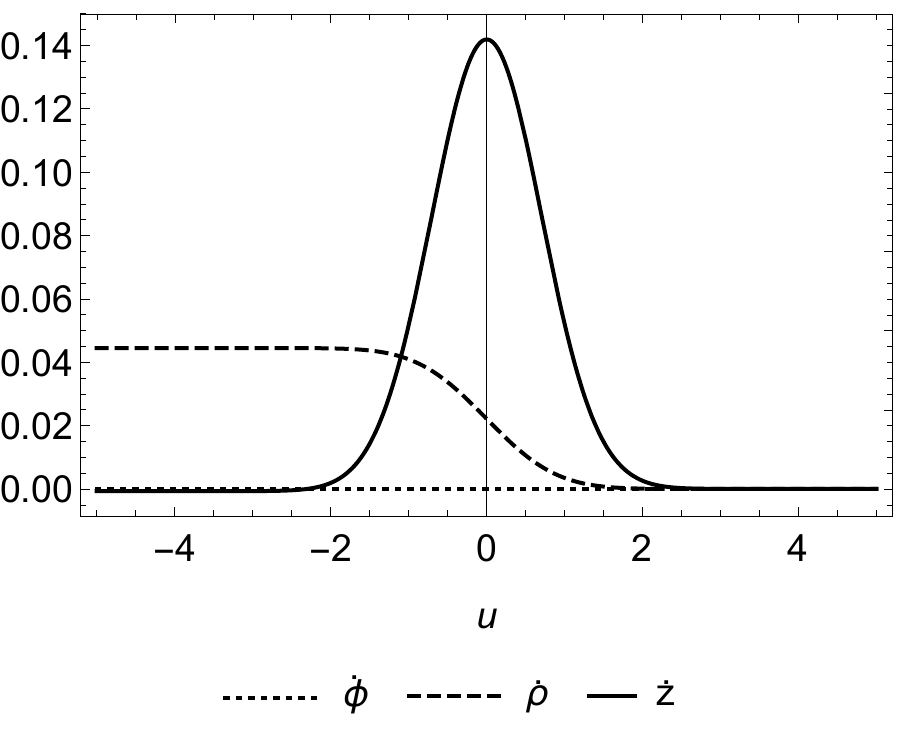}
	\caption{Velocity of a free particle initially at rest in $\rho (u\rightarrow\infty)=5$  with $J=0$.}
	\label{fig9}
\end{minipage}
\end{figure}

\begin{figure}[htbp]
\centering
\begin{minipage}[t]{.45\textwidth}
\centering
	\includegraphics[width=0.95\textwidth]{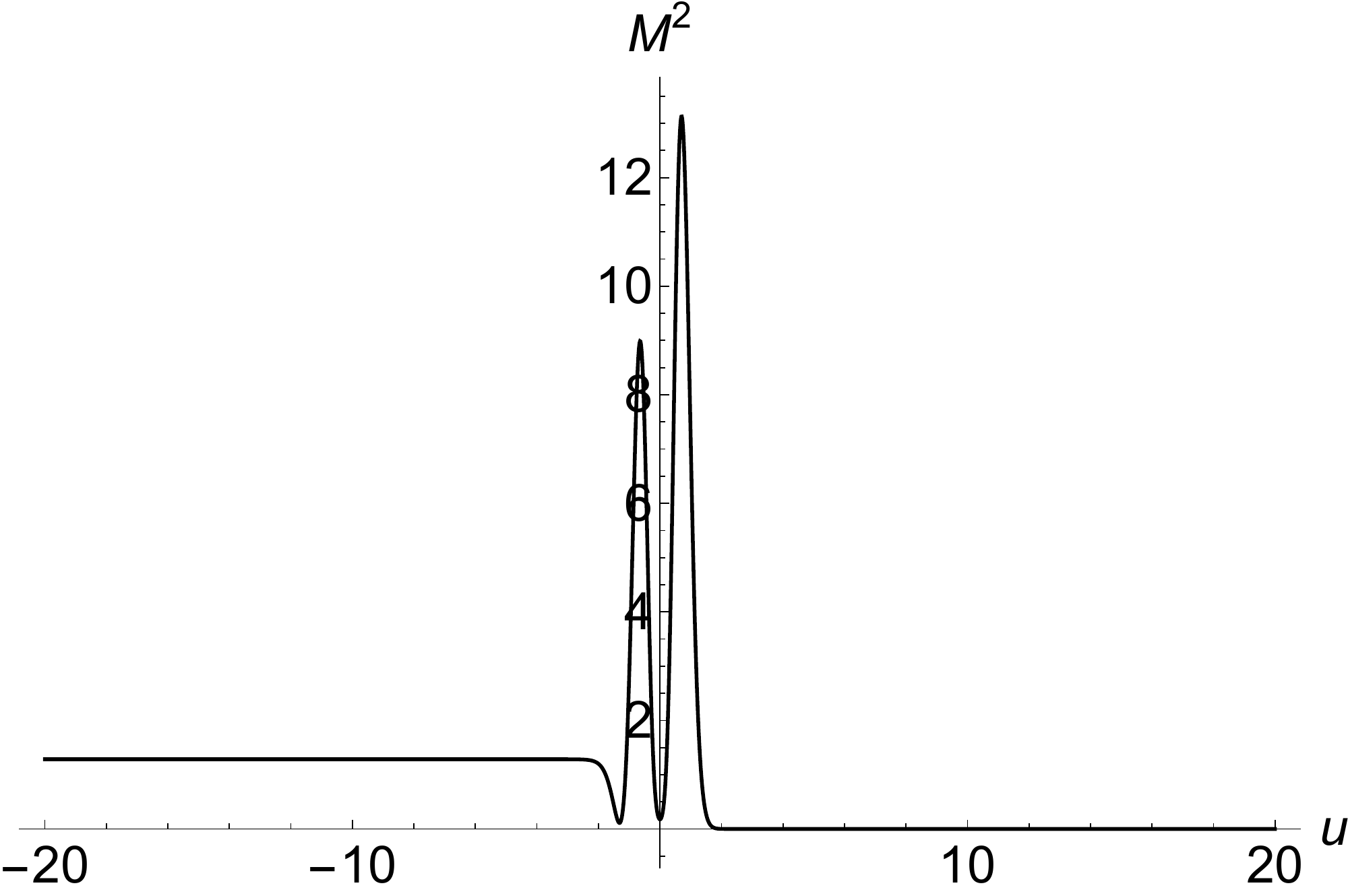}
	\caption{Components of the angular-momentum of a free particle initially at rest in $\rho (u\rightarrow\infty)=5$ with $J$ given by (\ref{eq43}).}
	\label{fig10}
\end{minipage}
\begin{minipage}[t]{.45\textwidth}
\centering
	\includegraphics[width=0.95\textwidth]{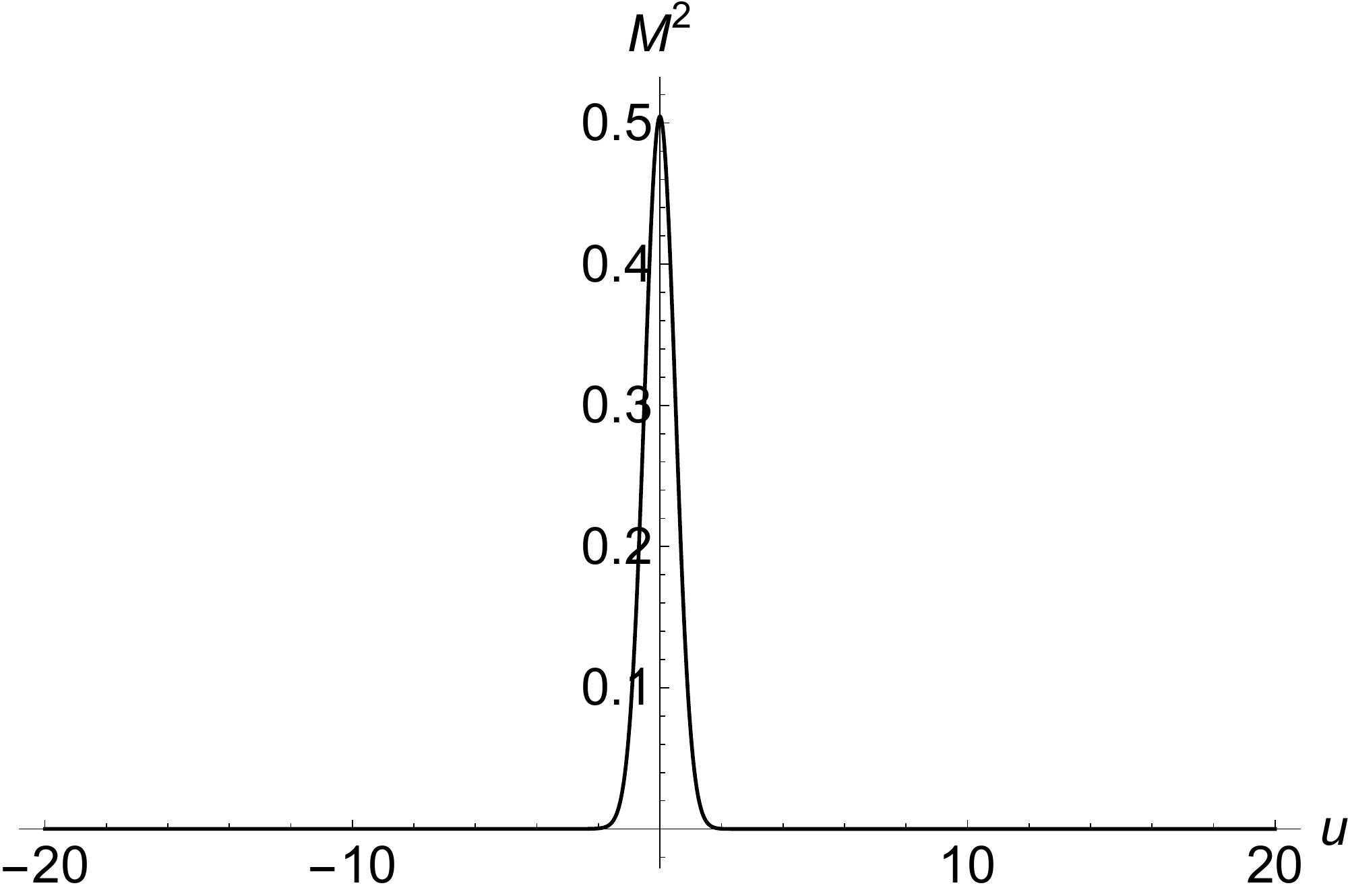}
\caption{Components fo the angular-momentum of a free particle initially at rest in $\rho (u\rightarrow\infty)=5$ with $J=0$.}
	\label{fig11}
\end{minipage}
\end{figure}

The presence of the gyratonic term significantly affects the radial angular-momentum density of the field. If the gravitational waves can be detected by their effects on particles, it is interesting to consider an example of how the gyratonic term affects some physical properties of free particles. Let us analyze the case of a particle, initially free of any forces, that is hit by an axially symmetric wave. The trajectory of the particle is obtained by numerically solving the geodesic equations (9-11) of the Ref. \cite{maluf2018kinetic}, parametrized with respect to $u$.
The effect of the gyratonic term may be better perceived by considering a wave solution with a more prominent gyratonic characteristic. Considering a wave with $J$ given by
\begin{equation}\label{eq43}
J=8\frac{d}{du}e^{-u^{2}}
\end{equation}
and $H$ given by equation (\ref{eq39}), a particle initially at rest at position $\rho (u\rightarrow\infty)=5$ achieve a three-dimensional motion. The same particle has a movement constrained into a plane when hit by a non gyratonic wave. In both cases, for the tested solution, the particle does not remain at rest after the passage of the wave, i.e., there is no permanent exchange of energy between the particle and the field \cite{Maluf2018}.
This can be best seen by evaluating the components of the velocity of the particle in both cases.
This is shown in the Fig.s \ref{fig8} and \ref{fig9}.
We can see in the first one that the gyratonic wave imparts a permanent displacement on the particle, i.e., after the passage of the wave the radial velocity tends to a non-null constant value. Note that by the definition (\ref{eq13}) a positive parameter $u$ indicates a negative time $t$, so the initial conditions represent a particle at rest before the wave passes. 
As presented in the expressions (\ref{eq29}) and (\ref{eq38}) the physical difference between axially symmetric gyratonic and non gyratonic pp-waves is the presence of a non-null radial angular-momentum density. This ensures that the different behaviors of the velocity in the Figs. \ref{fig8} and \ref{fig9} are due to the wave radial angular-momentum density only. The non-nullity of the radial angular-momentum density of the field affects the behavior of the particles by changing their angular-momentum components. To demonstrate this, we consider the angular-momentum per unit of mass for a classical particle, given by
\begin{equation}\label{eq44}
\vec{M}=\frac{1}{\sqrt{2}}\rho z \dot{\phi}\hat{\rho}+\frac{1}{\sqrt{2}}\left(\rho\dot{z}-z\dot{\rho}\right)\hat{\phi}-\frac{1}{\sqrt{2}}\rho^{2}\dot{\phi}\hat{z}\,.
\end{equation}
The total angular-momentum $M^{2}\equiv\vec{M}\cdot\vec{M}$ is permanently altered 
During the passage of the wave, i.e., the gyratonic term induces a permanent variation in the particle total angular-momentum $M^{2}\equiv\vec{M}\cdot\vec{M}$, as can be seen in Fig. \ref{fig10}. In the non gyratonic case, where the field has only azimuthal angular-momentum density, there is not permanent exchange of angular-momentum between the wave and the particle, as can be seen in Fig. \ref{fig11}.
We can then conclude that the angular-momentum of the field is directly connected to the angular-momentum of the particle, especially the radial angular-momentum density of the field.

A direct analysis between the angular-momentum of the particle and the angular-momentum of the field cannot be made in the coordinates used in this paper. An integration of the angular-momentum density of the gravitational field produces an expression that is a function of the variable $t$, while the geodesic equations of the particle are parametrized by the variable $u$ which is related with the time $t$ and the coordinate $z$. Therefore, only a qualitative analysis can be obtained. In order to have a consistent quantitative analysis, one must construct the tetrads associated with the line element in (\ref{eq12}) in pure Brinkmann coordinates, establishing the appropriated spatially static condition for this tetrads. This will be pursued elsewhere.


\end{document}